\documentstyle[preprint,revtex,eqsecnum]{aps}
\begin{document}
\draft
\begin{title}
On Models with Inverse-Square Exchange
\end{title}
\author{Z. N. C. Ha and F. D. M. Haldane}
\begin{instit}
Department of Physics, Princeton University, Princeton, New Jersey 08544
\end{instit}

\receipt{25 April, 1992}
\begin{abstract}
A one-dimensional quantum N-body system of either fermions or bosons
with $SU(n)$ ``spins'' (or colors in particle physics language) interacting
via inverse-square exchange is presented in this
article.  A class of eigenstates
of both the continuum and lattice version of the model Hamiltonians is
constructed in terms of the Jastrow-product type wave function.
The class of states we construct in this paper corresponds to the
ground state and the low energy excitations
of the model that can be described by the effective harmonic fluid Hamiltonian.
By expanding the energy about the ground state we find the
harmonic fluid parameters (i.e. the charge,
spin velocities, etc.), explicitly.
The correlation exponent and the compressibility are also found.
As expected
the general harmonic relation(i.e. $v_S=(v_Nv_J)^{1/2}$) is satisfied among
the charge and spin velocities.
\end{abstract}

\section{Introduction}
\label{sec:intro}
In this paper we report an exact treatment of a one-dimensional system
of particles with SU(n) ``spins'' interacting with the inverse-square, two-body
exchange in continuum space.  In particular, we show that a class of
eigenstates of the
Hamiltonian is given by the Jastrow-product type of wave function.
We also show that this model can be put on the lattice for all the
positive integer values of the dimensionless interaction parameter which is
defined in Eq. (\ref{firsteq}).
The continuum version is a direct generalization of Sutherland's model of
interacting spinless fermions (or hard core bosons)\cite{suther}.
The lattice version
corresponds to the $SU(n)$ t-J model which, at $n=1$, maps to the
Haldane-Shastry model \cite{hasha}.

Fig. 1 shows how various models with the inverse square exchange have
evolved from the original Sutherland's Model.  In the figure the models that
already exist are in rectangular boxes and the new models that we present in
this paper are in ovals.  The solid arrows mean ``generalized from'' or
``evolved from''.  Two equivalent models that are represented in different
ways are connected by a dashed arrow.  We emphasize that while the bosonic
t-J model ($\lambda$ odd) is a direct generalization of the bosonic
supersymmetric case ($\lambda = 1$), the fermionic t-J model ($\lambda$ even)
is not directly related to the fermionic supersymmetric t-J model that
Kuramoto and Yokoyama \cite{ky} found.

\section{Multicomponent Generalization of Sutherland's Model}
\label{sec: gen}
For convenience we consider a periodic system, where
the Hamiltonian in units of $\hbar^2/m$ is given by
\begin{equation}
H = -{1\over 2} \sum_i {\partial^2\over \partial x_i^2} + \lambda
\sum_{i<j} {\lambda + P^\sigma_{ij}\over d(x_i-x_j)^2}, \label{firsteq}
\end{equation}
where $\lambda$ is the dimensionless interaction parameter.
$P^\sigma_{ij}$ is an operator that exchanges particle spins at $x_i$ and
$x_j$, and $d(x) = (L/\pi)|\sin (\pi x/L)|$.
$d(x_i-x_j)$ is the chord distance between
particles at $x_i$ and $x_j$ on a circle with
circumference $L$.
If all particles have the same spin, this model reduces to the system of
spinless particles studied by Sutherland. Note that Sutherland's coupling
parameter $\lambda^\prime$ corresponds to $\lambda -1$ in our notation. For
example, the spinless free Fermi gas corresponds to $\lambda^\prime = 1$, but
in our notation $\lambda = 0$.

It is easily found that the wave function $\Psi$ must vanish as
$|x_i-x_j|^{\lambda+1}$ ($|x_i-x_j|^\lambda$) as $x_i-x_j\rightarrow 0$
in the case of symmetric (antisymmetric) spin configuration of the two
particles at $x_i$ and $x_j$.
And, if $0<\lambda <1$ the effective interaction strength is attractive for the
antisymmetric spin configuration  and there is some ambiguity in this case as
it must be further specified whether $\Psi$ vanishes as $|x_i-x_j|^\lambda$ or
as $|x_i-x_j|^{1-\lambda}$ as particles approach. If we choose the first
boundary condition the free fermion limit is obtained as $\lambda\rightarrow
0$.

There are two possible interpretations for this system.
The first is to consider the system as a ring embedded in a plane.
Hence, even though the particles are constrained to move in one-dimension, the
interaction is two-dimensional in nature.
The other is to regard the system as strictly one-dimensional by taking
$1/d(x_i-x_j)^2$ as the effective interaction after summing the
pairwise interaction of the particles around the ring infinite times.
The following identity,
\begin{equation}
\sum_{n=-\infty}^{+\infty} {1\over (x+nL)^2} = {1\over d(x)^2},
\end{equation}
has first been used by Sutherland \cite{suther} to show
the validity of the latter interpretation.

In analogy with the states previously constructed for the $SU(2)$ spin chain
in ref. \cite{hasha},
we propose the following Jastrow-product type wave functions
for our Hamiltonian:
\begin{equation}
\Psi(\{z\sigma\})=\Psi_0 \prod_{k}z_k^{J_{\sigma_k}},
\end{equation}
where
\begin{eqnarray}
\Psi_0 &=& \prod_{n>m}\phi_{nm} \nonumber \\
\phi_{nm} &=& |z_n-z_m|^{\lambda-x} (z_n-z_m)^{x+\delta_{\sigma_n\sigma_m}}
\exp(i{1\over 2}\pi \mbox{sgn}(\sigma_n-\sigma_m)).
\end{eqnarray}
Here, $\delta$ is the Kronecker delta function and $z_n = \exp(2\pi i x_n/L)$.
$\sigma_n$ is the ordered spin index and
$J_\sigma$ the global current of particles with $\sigma$ spin.
We take $J_\sigma$ to be an integer and will discuss the restrictions on the
allowed values of $J_\sigma$ later in this article. Note also that the wave
function with $\lambda=x=0$ is the Slater determinant that corresponds to the
states of free $SU(n)$ fermions.

The symmetry of the wave function with respect
to the exchange of particles is given as
\begin{equation}
\Psi(\ldots,z_i\sigma_i,\ldots,z_j\sigma_j,\ldots)=
(-1)^{x+1}\Psi(\ldots,z_j\sigma_j,\ldots,z_i\sigma_i,\ldots).
\end{equation}
Hence, for bosons(fermions), $x=1(x=0)$.

We write the total Hamiltonian as $H=H^0 + H^1 + H^2$ where $H^0$, $H^1$ and
$H^2$ are the
kinetic, potential and spin exchange Hamiltonian, respectively.
We will show that each operator acting on the wave function
gives two types of terms, ``wanted'' and ``unwanted''.  ``Wanted'' terms are
defined to be the terms that depend only on the global variables such as
the total number of particles, $J_\sigma$, etc.  ``Unwanted'' terms explicitly
depend on the local variables such as $z_i$.
Since the eigenenergy should depend only on the global variables, the
``unwanted'' terms for $H^0$, $H^1$ and $H^2$ should cancel or combine to
give ``wanted'' terms.

We first examine $H^0$ acting on the wave function.
We define the following derivatives:
\begin{eqnarray}
\varphi_{ij}&\equiv&{\partial_{z_j}\phi_{ij}\over \phi_{ij}}=
-{\lambda\over z_i-z_j}-{\lambda-x\over 2z_j}-{\delta_{\sigma_i\sigma_j}\over
z_i-z_j}  \\
\xi_{ij}&\equiv&\partial_{z_j}\varphi_{ij} \\
\eta^{(1)}_j&\equiv&z_j{\partial_{z_j}\Psi_0\over \Psi_0}=z_j\sum_{i(\ne
j)}\varphi_{ij} \\
\eta^{(2)}_j&\equiv&z^2_j{\partial_{z_j}^2\Psi_0\over \Psi_0}=z^2_j\sum_{i(\ne
j)}\xi_{ij}+ (\eta^{(1)}_j)^2.
\end{eqnarray}
In terms of these derivatives we find that $H^0\Psi = 2(\pi/L)^2
\sum_j[\eta^{(2)}_j+(2J_{\sigma_j}+1)\eta^{(1)}_j+J^2_{\sigma_j}]\Psi$.
After some algebra one can show that
$H^0$ acting on the wave function gives
``wanted''($W_K$) and ``unwanted''($U_K$) terms. We write these terms as
follow:
\begin{equation}
H^0\Psi = 2\left(\pi\over L\right)^2(W_K + U_K)\Psi,
\end{equation}
where
\begin{eqnarray}
W_K &=&{\lambda^2\over 12}N(N^2-1)+{1\over 2}(\lambda+1)\sum_{\sigma}M_\sigma
(M_\sigma-1)+{1\over 3}\sum_{\sigma}M_\sigma(M_\sigma-1)(M_\sigma-2)
\nonumber \\
&-&\sum_{\sigma}M_\sigma J_\sigma(\lambda N - M_\sigma - J_\sigma - \lambda
+1) + E(x), \\
U_K &=& \lambda\sum_{i\ne j}{z_iz_j\over (z_i-z_j)^2}\left\{\lambda - 1 +
2\delta_{\sigma_i\sigma_j}\right\} +\lambda\sum_{i\ne j}{z_iJ_{\sigma_i}-
z_jJ_{\sigma_j}\over z_i-z_j} \nonumber \\
&+&\lambda\sum_{(i\ne j\ne k)}{z_i(z_i+z_k)
\over (z_i-z_j)(z_i-z_k)}\delta_{\sigma_i\sigma_j},
\end{eqnarray}
where $E(x)=x^2 N(N-1)^2/4 + (x/2)(N-1)\sum_{\sigma}M_\sigma
(M_\sigma-1)+x(N-1)\sum_{\sigma}M_\sigma J_\sigma$. $M_\sigma$ is the
total number of $\sigma$ particles in the system.
We can now anticipate $U_K$ to be canceled or combined with the ``unwanted''
terms from $H^1$ and $H^2$ to give local variable independent terms.

$H^1$ acts trivially on $\Psi$ to give $\lambda^2\sum_{i<j}1/d(x_i-x_j)^2$.
The action of $H^2$ on the
wave function is less trivial. We need to evaluate the following expression:
\begin{equation}
P_{\lbrace M \rbrace}\equiv{1\over 2 \Psi}\sum_{i>j}
{P^\sigma_{ij} \Psi \over d(x_i-x_j)^2}= -{1\over \Psi(\{z\sigma\})}
{\sum_{i\ne j}{z_i z_j \over (z_i-z_j)^2}\Psi(\{ z
\sigma^\prime \})},
\end{equation}
where $\{ z\sigma^\prime \}$ is a
configuration with $\sigma_i$ and $\sigma_j$ exchanged with
respect to $\{z\sigma\}$.  Similarly, we let $\{z^\prime\sigma\}$ be equal to
$\{z\sigma\}$ with $z_i$ and $z_j$ exchanged. Using the identity,
$\Psi(\{ z\sigma\}) = (-1)^{x+1}\Psi(\{ z^\prime\sigma^\prime\})$, we may
then rewrite the exchange operation as
\begin{eqnarray}
P_{\{M\}}&=&(-1)^{x+1}\sum_{i\ne j}{z_i z_j\over (z_i-z_j)^2}
{\Psi(\{z\sigma^\prime\})
\over \Psi(\{ z^\prime\sigma^\prime\})}\nonumber \\
&=&\sum_{i\ne j}{z_i z_j\over (z_i-z_j)^2}(-1)^{\delta_{\sigma_i\sigma_j}}
{\left( z_i \over z_j\right)}^{J_{\sigma_j}-J_{\sigma_i}}\prod_{k\ne ij}
\left( z_k-z_i \over z_k-z_j \right)^
{\delta_{\sigma_j\sigma_k} - \delta_{\sigma_i\sigma_k}}
\label{eneg}
\end{eqnarray}
The action of $P^\sigma$ is same for both boson and fermion cases
and is independent of the interaction parameter $\lambda$.

We use the following identity,
\begin{eqnarray}
{\left( z\over z^\prime \right)}^n\equiv \sum_{q=1}^{|n|}{|n|\choose q}\left\{
{\left( z-z^\prime \over z^\prime \right)}^q \theta(n)
+{\left( z^\prime-z \over z \right)}^q \theta(-n)\right\},\label{jj}
\end{eqnarray}
to rewrite Eq.\ (\ref{eneg}) as
\begin{equation}
P_{\{M\}} = P_0 + \sum_{\sigma,\sigma^\prime}\sum_{q=1}^
{|J_{\sigma^\prime}-J_\sigma|}
 P_q^{\sigma\sigma^\prime}.\label{em}
\end{equation}
$\theta$ in Eq.\ (\ref{jj}) is the step function with $\theta(0)=1/2$, and
$P_0$ and $P_q^{\sigma\sigma^\prime}$ are given by
\begin{eqnarray}
P_0&=&-\sum_{i\ne j}{z_i z_j\over (z_i-z_j)^2}\left\{\delta_{\sigma_i\sigma_j}
+ (1-\delta_{\sigma_i\sigma_j})
\prod_{k\ne ij}\left({z_k-z_i\over
z_k-z_j}\right)^{\delta_{\sigma_j\sigma_k}-\delta_{\sigma_i\sigma_k}}\right\}\\
P_q^{\sigma\sigma^\prime}&=&2\sum_{i\ne j}{z_i z_j\over (z_i-z_j)^2}
\delta_{\sigma\sigma_i}\delta_{\sigma^\prime\sigma_j}
(1-\delta_{\sigma\sigma^\prime}) \prod_{k\ne ij}\left({z_k-z_i\over
z_k-z_j}\right)^{\delta_{\sigma_j\sigma_k}-\delta_{\sigma_i\sigma_k}}
\nonumber\\
&\times&{J_{\sigma^\prime}-J_\sigma\choose q}\theta(J_{\sigma^\prime}-J_\sigma)
\left(z_i-z_j\over z_j\right)^q.
\end{eqnarray}

It is useful to separately consider the terms with $q=0,1$
and the terms with $q\ge 2$.
For the terms with $q=0,1$
we introduce two sets of site indices $\{\alpha \}$ and
$\{\beta \}$.  The set $\{\alpha \}$ ($\{\beta \}$) includes all the
locations of particles with the spin $\sigma$($\sigma^\prime$).
Using the identity,
\begin{equation}
\left(z_k-z_i\over z_k-z_j\right)^{
\delta_{\sigma^\prime\sigma_k}-\delta_{\sigma\sigma_k}}
\equiv 1-\delta_{\sigma^\prime\sigma_k}{z_i-z_j\over z_k-z_i}
+\delta_{\sigma\sigma_k}{z_i-z_j\over z_k-z_j}\;\;
\mbox{for}\; \sigma\ne \sigma^\prime,
\end{equation}
the products in $P_{\{M\}}$ may be expanded and
the typical terms in $P_{\{M\}}$ can be simplified using the following two
theorems.

{\it Theorem 1\/}:
Let $\{\alpha\}$ and $\{\beta\}$ be two sets of distinct integers between $1$
and $N$, and let $0\le q\le 1$ and
$\Delta=(1-\delta_{\sigma,{\sigma^\prime}})\delta_{\sigma\sigma_i}
\delta_{\sigma\sigma_{\alpha_1}}\!\!\!\cdots \delta_{\sigma\sigma_{\alpha_n}}
\delta_{\sigma^\prime\sigma_j}
\delta_{\sigma^\prime\sigma_{\beta_1}}\!\!\!\cdots
\delta_{\sigma^\prime\sigma_{\beta_m}}.$
Then,
\begin{eqnarray}
\sum_{n=1}^{M_\sigma-1}&{}&\!\!\!\sum_{m=1-q}^{M_{\sigma^\prime}-1}\sum_{i\ne
j}\sum_{\{\alpha\},\{\beta\}}{(-1)^m\over n!\,m!}{z_i z_j^{1-q}
(z_i-z_j)^{n+m-2+q}\over (z_{\alpha_1}-z_i)\cdots(z_{\alpha_n}-z_i)
(z_{\beta_1}-z_j)\cdots(z_{\beta_m}-z_j)}\Delta \nonumber \\
&=&\left\{
\begin{array}{l}
-\sum_{k=1}^{Min(M_\sigma,M_{\sigma^\prime})}(M_\sigma-k)(M_{\sigma^\prime}-k)
 \;\;\mbox{for}\; q=0\;,\\
-\sum_{k=1}^{Min(M_\sigma,M_{\sigma^\prime})}(M_\sigma-k)\;\; \mbox{for}\;
q=1\;.
\end{array}\right.\nonumber 
\end{eqnarray}

{\it Theorem 2\/}:
For $q\ge 2$ the following identity will hold for $M_\sigma \ge
M_{\sigma^\prime}$,
\begin{eqnarray}
\sum_{i\ne j}&{}&\!\!\!{z_i\over z_j}\left({z_i-z_j\over z_j}\right)^{q-2}\!\!
(1-\delta_{\sigma{\sigma^\prime}})
\delta_{\sigma\sigma_i}\delta_{\sigma^\prime\sigma_j}
\prod_{k\ne ij}\left({z_k-z_i\over z_k-z_j}\right)^
{\delta_{\sigma^\prime\sigma_k} - \delta_{\sigma\sigma_k}} \nonumber \\
&=&\left\{
\begin{array}{l}
M_{\sigma^\prime}\;\; \mbox{for}\; q=2,\\
0 \;\;\mbox{for} \;2 < q \le M_\sigma-M_{\sigma^\prime}+1.
\end{array}\right. \nonumber 
\end{eqnarray}
Proofs of these theorems are given in Appendix A and B.

After reorganizing some terms we obtain the following results:
\begin{equation}
P_{\{M\}} = W_P + U_P, \label{suneng}
\end{equation}
where
\begin{eqnarray}
W_P&=&-{1\over 3}\sum_\sigma M_\sigma (M_\sigma -1)(M_\sigma -2)
-{1\over 3}\sum_{\sigma < \sigma^\prime} M_{\sigma^\prime}(M_{\sigma^\prime} -
1)(3M_\sigma-M_{\sigma^\prime}-1), \nonumber\\
&-&\sum_{\sigma < \sigma^\prime} M_{\sigma^\prime}(M_\sigma -
M_{\sigma^\prime}) (J_{\sigma^\prime} - J_{\sigma})
+\sum_{\sigma < \sigma^\prime}
M_{\sigma^\prime}(J_{\sigma^\prime}-J_\sigma)^2, \label{WP}\\
U_P&=&\sum_{i\ne j}{z_iz_j\over (z_i-z_j)^2}\left\{1 -
2\delta_{\sigma_i\sigma_j}\right\} +\sum_{i\ne j}{z_iJ_{\sigma_j}-
z_iJ_{\sigma_i}\over z_i-z_j} \nonumber \\
&-&2\sum_{(i\ne j\ne k)}{z_iz_j
\over (z_i-z_j)(z_i-z_k)}\delta_{\sigma_i\sigma_k}.\label{unwanted}
\end{eqnarray}
Because of symmetry we can choose $M_1\ge M_2 \ge\ldots\ge M_n$ and
$0\le J_{\sigma^\prime}-J_\sigma\le M_\sigma - M_{
\sigma^\prime}+1$ for $\sigma^\prime > \sigma$ without loss of generality.
And, one can easily check that $U_K,U_P$ and the potential energy term combine
to give local-variable independent terms.

Before we give the full expressions for the eigenenergy, we discuss the allowed
values
of the integer currents.
{\it Theorem 2} gives a simple rule for selecting allowed currents.
If we choose $-1\le J_{\sigma^\prime}-J_\sigma \le
M_\sigma-M_{\sigma^\prime}+1$ for $M_\sigma \ge M_{\sigma^\prime}$,
then $P_q = 0$ for $q>2$.  Otherwise, the energy will not in general be
independent of the local variables and the corresponding wave function will
not be an eigenfunction.
Pictorial illustration of the allowed
currents is shown in Fig.~\ref{curr}.
Each row of $M_\nu$ boxes represents $M_\nu$ ``particles'' of same spin.
A single box gives a unit of current.
Fig.~\ref{curr}(a) is the reference state with no current.
If the current is $j > 0 (< 0)$ for a row, the row is moved $j$ boxes to the
right(left) as shown in Fig.~\ref{curr}(b). In order for a state to be an
eigenstate, the following condition must hold true: {\it
For a given pair of rows of boxes all except the first and the last boxes in
the row with smaller number of boxes must be positioned within the other row.}
Fig.~\ref{curr}(c), for example, cannot be an eigenstate because
the last two boxes in the second row are not within the first row.

Finally, the eigenenergies of the model can be given
as follow:
\begin{equation}
E={2\hbar^2 \pi^2\over m L^2}(E_1 + E_2 + E(x)),
\end{equation}
where $E_1$ and $E_2$ are energies due to one- and two-spin interactions,
respectively, and
are given by
\begin{eqnarray}
E_1&=&{\lambda^2\over 12}N(N^2-1) +{1\over 2}\lambda N\sum_\sigma M_\sigma(
M_\sigma-1)+{1\over 6}(1-\lambda)\sum_\sigma M_\sigma(M_\sigma-1)(2M_\sigma-1)
\nonumber \\
&+&\sum_\sigma J_\sigma M_\sigma(M_\sigma+J_\sigma-1) \\
E_2&=&-{1\over 3}\lambda\sum_{\sigma<\sigma^\prime}
M_{\sigma^\prime}(M_{\sigma^\prime}-1)(3M_\sigma-M_{\sigma^\prime}-1)
-\lambda\sum_{\sigma<\sigma^\prime}M_{\sigma^\prime}
(M_\sigma-M_{\sigma^\prime})(J_{\sigma^\prime}-J_\sigma)\nonumber \\
&+&\lambda\sum_{\sigma<\sigma^\prime}M_{\sigma^\prime}
(J_{\sigma^\prime}-J_\sigma)^2 \label{E2}
\end{eqnarray}
The ground state for fermions(bosons) is given by the following two conditions:
$(i) M_{\sigma^\prime} = M_\sigma = M$ and $(ii) J_{\sigma^\prime}=
J_\sigma = J= -(M-1)/2,\;\;
-(N+M-1)/2 $ for fermions and bosons, respectively.  The currents should be
integers. If the condition $(ii)$ does not give an integer current the ground
state is degenerate and the ground state currents are two integers closest to
the
the half odd integer value.

\section{SU(n) t-J Model}
In ref. \cite{hasha}, Sutherland's model for a system of spinless particles
in continuum space was extended to the lattice.  Using the similar procedures
we put our model on the lattice.  We propose the following Hamiltonian:
\begin{equation}
H= {\cal P}\left(\sum_{j=1}^{N}\sum_{n=1}^{N_a-1} {t^n_j \over d(n)^2} +
\sum_{i > j}
{l(l + P^{\sigma}_{ij})n_i n_j\over d(x_i-x_j)^2}\right)
{\cal P}. \label{lhamil}
\end{equation}
$l$ is a positive integer.
$N_a$ is the total number of sites and N the total number of particles.
$d(n) = (N_a/\pi)|\sin (n)|$.
$t_j^n$ hops a particle at $j$th site to $(j+n)$th (mod $N_a$) site.
$n_j$ is equal to one(zero) if $j$th site is occupied (empty).
$P^\sigma_{ij}$ is, as before, the spin exchange operator.
${\cal P}$ is the projection operator that insures the absence of
multiply occupied sites.

In analogy with ref. \cite{hasha} we propose the following eigenstate for the
model Hamiltonian,
\begin{equation}
|\Psi_{\lbrace M \rbrace}\rangle = \sum_{\lbrace z\sigma \rbrace}
{\phi(\lbrace z\sigma \rbrace)} | \lbrace z\sigma \rbrace
\rangle.\label{wf}
\end{equation}
Since the total number of each spin is a good quantum number we classify
the eigenstates into sectors labeled by
$\lbrace M \rbrace \equiv (M_1, M_2,\ldots, M_n)$, where
$M_\nu$ is the number of $\nu$th spin and $\sum_{\nu = 1}^n M_\nu = N$. We also
represent the particle
configuration as $\lbrace z\sigma \rbrace \equiv
(z_1\sigma_1,\ldots,z_i\sigma_i,
\ldots,z_j\sigma_j,\ldots, z_N\sigma_N)$, where
$z_i$ and $\sigma_i$ are the location and spin of the $i$th particle.
The sum in Eq.\ (\ref{wf}) is over $N_a!/(M_h!M_1!\,\cdots\,M_n!)$ distinct
spin configurations of the sector $\{M\}$ where $N_a$ and $M_h$ are the
total number of sites and the number of holes, respectively.
The function $\phi$ in
Eq.\ (\ref{wf}) is given by the following Jastrow-product,
\begin{equation}
\phi(\lbrace z \rbrace | \lbrace \sigma \rbrace) = \prod_{i<j}(z_i - z_j)
^{l+\delta_{\sigma_i\sigma_j}} e^{i{\pi \over 2}sgn(\sigma_i - \sigma_j)}
\prod_k {z_k^{J_{\sigma_k}}}.
\end{equation}
Considering the symmetry of the wave function we take odd (even) $l$
to be bosonic (fermionic) state.

As in the continuum case we break up the total Hamiltonian into the kinetic,
potential, and spin exchange parts ($H^0_L$, $H^1_L$ and $H^3_L$,
respectively).
The actions of $H^1_L$ and $H^3_L$ are same as
the corresponding operators in the continuum case.
For $H^0_L$ we
have the following relations:
\begin{eqnarray}
<H^0>&\equiv&{{\langle z_1\sigma_1,\ldots,z_N
\sigma_N | H^0 | \Psi_
{\lbrace M \rbrace} \rangle} \over {\langle z_1\sigma_1,\ldots,
z_N\sigma_N | \Psi_{\lbrace M \rbrace} \rangle}} \nonumber \\
&=&4{\sum_{i=1}^{N}\sum_{n=1}^{N_a-1}z^{nJ_{\sigma_i}}(1-z^n)^{-1}
(1-z^{-n})^{-1}\prod_{j(\ne i)}\left(z_iz^n-z_j\over z_i-z_j\right)^{l+
\delta_{\sigma_i\sigma_j}}} \label{lateng}
\end{eqnarray}
where $|z_1\sigma_1,\ldots,z_N\sigma_N\rangle$ is one of the basis states
of the sector $|\Psi_{\{M\}}\rangle$.
In order to evaluate $<H^0_L>$ we need the following theorem (a slightly
different version of this theorem can be
found in ref. \cite{hasha}).

{\it Theorem 3\/}:
Let $J$ and $p$ be non-negative integers and $z=exp(2\pi i/N_a)$. Then,
\begin{eqnarray}
\sum^{N_a-1}_{n=1}z^{nJ}(1-z^{-n})^{-1}(1-z^n)^{p-1}&=&
{N_a^2-1\over 12}-{J(N_a-J)\over 2}\;\;\mbox{for}\;p=0,\;\nonumber\\
&=&-J+{N_a-1\over 2}\;\;\mbox{for}\;p=1,\;\nonumber\\
&=&+1\;\;\mbox{for}\;p=2,\;0\le J\le N_a-2\nonumber \\
&=&-(N_a-1)\;\;\mbox{for}\; p=2,\;J=N_a-1\nonumber\\
&=&\sum_{m>0}(-1)^{mN_a-J}{p-2\choose mN_a-J-1}N_a\;\;
\mbox{for}\;p\ge 3,\;\nonumber
\end{eqnarray}
m is a positive integer, and
the restriction on the current is $0\le J \le N_a-1$ unless
specified otherwise.

If we take $l>0$,
we can multiply out the product in Eq. (\ref{lateng}).  The resulting
expression will be a sum of
terms containing a factor $(1-z^n)^p$.
Since the maximum value of $p$ is $l(N-1)+M_\sigma-1$,
if we choose
$0\le J_\sigma \le N_a-l(N-1)-M_\sigma+1$, we may ignore terms with
$p\ge 3$.
We combine the remaining terms with those from $H^1_L$ and $H^2_L$
and obtain the following eigenenergies for the lattice model:
\begin{equation}
E^L = E^L_1 + E^L_2
\end{equation}
where
\begin{eqnarray}
E^L_1 &=& {1\over 6}N(N_a^2-1)+{l^2\over3}N(N-1)(N-2)-{l\over2}
(N_a-l)N(N-1) \nonumber \\
&-&\left({N_a-1\over 2}-l(N-1)\right)\sum_{\sigma}
M_\sigma(M_\sigma-1) + {1\over 3}(1-l)\sum_{\sigma}M_\sigma(M_\sigma-1)
(M_\sigma-2) \nonumber \\
&-&\sum_{\sigma}M_\sigma J_\sigma(N_a-l N- M_\sigma -J_\sigma
+l+1), \\
E^L_2&=& E_2 ,
\end{eqnarray}
where $E_2$ is given by Eq. (\ref{E2}).
We emphasize at this point that the above expressions for the eigenenergies
for the lattice model are valid only for $l N+M_\sigma\le N_a+l+1$.

For $l=1$ (bosonic supersymmetric case) we can find another expression
for the energy for $N+M_\sigma \ge N_a + 2$.  We proceed by rewriting
the kinetic energy term as
\begin{equation}
<H^0> = 2\sum_\sigma\sum_{i\ne j} (1-\delta_{\sigma\sigma_h})
\delta_{\sigma\sigma_i}\delta_{\sigma_h\sigma_j}\left({z_i\over z_j}
\right)^{J_\sigma+1} {z_iz_j\over (z_i-z_j)^2}\prod_{k\ne ij}
\left({z_k-z_i\over z_k-z_j}\right)^{\delta_{\sigma_h\sigma_k}-
\delta_{\sigma\sigma_k}},
\end{equation}
where $\sigma_h$ represents the hole.  The above expression is
almost identical with the expression for the spin exchange operator and
can easily be evaluated using {\it Theorem 1} and {\it 2}. The energy
then is given by
\begin{equation}
E(l=1) = {\pi^2\over 2N_a^2}(E_k(l=1) + W_P),\label{lone}
\end{equation}
where
\begin{eqnarray}
E_k(l&=&1)=-{M_h(N_a^2-1)\over 6} -{1\over 3}M_h(M_h-1)(M_h-2)
+M_h\sum_\sigma J_\sigma(J_\sigma-M_\sigma-1)
\nonumber \\
&+&{N_a-1\over 2}\left(\sum_\sigma M_\sigma(M_\sigma-1)+M_h(M_h-1)\right)
-{1\over 3}M_h(M_h-1)[3N-n(M_h+1)]
\nonumber \\
&+&M_hN + \sum_\sigma(J_\sigma-1)[2M_\sigma M_h - M_h(M_h+1)],
\end{eqnarray}
where the currents are restricted to $-(M_\sigma-M_h)\le J_\sigma\le 0$.
$W_P$ in Eq. (\ref{lone}) is given by Eq. (\ref{WP}).

For $l=0$ the wave function in Eq. (\ref{wf}) is that of
Gutzwiller projected free fermions and
no
longer an eigenstate of the Hamiltonian in Eq. (\ref{lhamil}).  However,
the wave function is  an eigenstate of the following Hamiltonian
\begin{equation}
H={\cal P}\left(-\sum_{j,n}{t^n_j\over d(n)^2}+
\sum_{i>j}{(P^\sigma_{ij}-1)n_in_j\over d(x_i-x_j)^2}\right){\cal P}.
\end{equation}
This Hamiltonian is SU(n) generalization of the SU(2) fermionic
supersymmetric t-J model which Kuramoto and Yokoyama \cite{ky} solved.
If we employ the similar technique used for $l=1$ case we may find the
eigenenergy given by Eq. (\ref{lone}) with $E_k(l=1)$ replaced with
$E_k(l=0)$ which is given by
\begin{equation}
E_k(l=0) = -{N(N_a^2-1)\over 6} + {1\over 2}(N_a-1)\sum_\sigma
M_\sigma(M_\sigma-1) + M_h\sum_\sigma J_\sigma(M_\sigma + J_\sigma -1).
\end{equation}
The currents are integers restricted to $-M_\sigma \le J_\sigma \le 0$.
The ground state energy is obtained if $J_\sigma = -(M_\sigma-1)/2$
and $M_\sigma$ odd. If $M_\sigma$ are not all odd,
the ground state is degenerate.

\section{SU(n) Spin Chain}
If we take the spin exchange Hamiltonian as our full Hamiltonian for a system
of spins on a lattice ring,
we obtain the $SU(n)$ spin chain model.  This is equivalent to taking the
``half filling'' limit
($N\rightarrow N_a$) of the lattice t-J model.  In this model the eigenenergies
for the boson and fermion systems are identical.  Since every site is occupied
with a spin, $U_P$ in Eq. (\ref{unwanted}) can now be summed and is equal to
$-N_a(N_a^2-1)/12 + \{(N_a-1)/2\} \sum_\sigma M_\sigma(M_\sigma-1)$. Hence,
$U_P$
is now local variable independent, and the eigenenergy is given by
Eq. (\ref{suneng}).

\section{The Harmonic Fluid Description}
We now turn our attention to the low energy excitation spectrum of our model
in the thermodynamic limit.
The low energy excitation spectrum of a one-dimensional
quantum fluid may be described by a sum of
two independent harmonic fluid Hamiltonians,
one for the charge and the other for the
spin.  This is a slight generalization of the spinless particle system in ref.
\cite{duncan}.  In general the effective Hamiltonian can be written as follow:
\begin{equation}
H=\sum_{\sigma,\sigma^\prime}\int dx \left( A_{\sigma\sigma^\prime}
\Pi_\sigma(x)\Pi_{\sigma^\prime}(x) + B_{\sigma\sigma^\prime}
\nabla\phi_\sigma(x)\nabla\phi_{\sigma^\prime}(x) \right)
\end{equation}
$\Pi_\sigma(x)$ is the local density fluctuation field as in ref.
\cite{duncan}.
$\phi_\sigma(x)$ is the canonical conjugate field to $\Pi(x)$,
$[\phi_\sigma(x),\Pi_{\sigma^\prime}(x^\prime)]=i\delta_{\sigma\sigma^\prime}
\delta(x-x^\prime)$.  Because of the $SU(n)$ symmetry we have
$A_{\sigma\sigma^\prime}=a_1+a_0\delta_{\sigma\sigma^\prime}$ and
$B_{\sigma\sigma^\prime}=b_1+b_0\delta_{\sigma\sigma^\prime}$. We
now express the Hamiltonian in terms of the Fourier transformed fields,
$\Pi_{\sigma k}$ and $\phi_{\sigma k}$.
\begin{equation}
H=\sum_k\sum_{\sigma,\sigma^\prime} A_{\sigma\sigma^\prime}\Pi_{\sigma k}
\Pi_{\sigma^\prime -k}+B_{\sigma,\sigma^\prime}k^2\phi_{\sigma k}
\phi_{\sigma^\prime -k}.
\end{equation}
We now construct normal mode fields as $\Pi^\nu_k = \sum_\sigma a^\nu_\sigma
\Pi_{\sigma k}$ and $\phi^\nu_k = \sum_\sigma b^\nu_\sigma\phi_{\sigma k}$.
If we choose $\sum_\sigma b^\nu_\sigma a^{\nu^\prime}_\sigma = \delta_
{\nu\nu^\prime}$, then $[\phi^\nu_k,\Pi^{\nu^\prime}_{-k^\prime}]=i\delta_
{kk^\prime}\delta_{\nu\nu^\prime}$.  We have the following equations
of motion: $[H,[H,\Pi^\nu_k]]=-(v^\nu k)^2 \Pi^\nu_k,\;\;
[H,[H,\phi^\nu_k]]=-(v^\nu k)^2 \phi^\nu_k$.
For $\Pi$-field we obtain the following equation,
\begin{equation}
(a_0b_0-(v^\nu)^2)a^\nu_\sigma + (a_1b_0 + a_0 b_1 + na_1b_1)\sum_\beta
a^\nu_\beta = 0
\end{equation}
The same equation with $a^\nu_\sigma$ replaced with
$b^\nu_\sigma$ holds for $\phi$-field.
There are only two possible values for $(v^\nu)^2$: $a_0b_0$ and
$(a_0+na_1)(b_0+nb_1)$. The first value corresponds to the case
$\sum_\beta a^\nu_\beta = 0$
and the second $\sum_\beta a^\nu_\beta \ne 0$.
Hence, the first would be the spin velocity and the
second the charge velocity.  The Hamiltonian can now be written as
\begin{equation}
H=\sum_k \left[ (a_0+na_1) \Pi^c_k\Pi^c_{-k} + (b_0+nb_1)
k^2\phi^c_k\phi^c_{-k} +
\sum_{s=1}^{n-1}\left\{a_0\Pi^s_k\Pi^s_{-k}+b_0k^2\phi^s_k\phi^s_{-k}
\right\}\right]
\end{equation}
We have one charge mode and $n-1$ spin modes; and because of the
$SU(n)$ symmetry all the spin modes have same velocity. We define the
following velocities associated with $\Pi$- and
$\phi$-fields for the charge(denoted as c) and spin(denoted as s) modes:
$v^c_N = a_0 + na_1,\; v^c_J = b_0+nb_1,\;
v^s_N = a_0,\; v^s_J = b_0,\; $.  The charge and spin velocities then are
$v_c=(v^c_N v^c_J)^{1/2}$ and $\;v_s=(v^s_N v^s_J)^{1/2}$.

Since the eigenstates for our model corresponds
to the low energy excitations of the model, we can easily obtain the harmonic
fluid parameters by examining the
energy expanded about the ground state.
By Taylor expansion we obtain the energy for the continuum model as follow,
\begin{equation}
E=e_0N + \mu N + {\pi\hbar\rho_0\over m}{\pi \hbar\over L}
\sum_{\sigma\sigma^\prime}{1\over 2}A_{\sigma\sigma^\prime}\Delta M_\sigma
\Delta M_{\sigma^\prime} + 2B_{\sigma\sigma^\prime}\Delta J_\sigma
\Delta J_{\sigma^\prime} \label{lowe}
\end{equation}
where
\begin{eqnarray}
e_0&=&{(\pi\hbar\rho_0)^2\over 6m}\left(\lambda+{1\over n}\right)^2,\\
\mu&=&{(\pi\hbar\rho_0)^2\over 2m}\left(\lambda+{1\over n}\right)^2,\\
A_{\sigma\sigma^\prime}&=&\lambda\left(\lambda+{1\over n}
\right)+\left(\lambda+{1\over n}\right)\delta_{\sigma\sigma^\prime},\\
B_{\sigma\sigma^\prime}&=&-{\lambda\over n}+
\left(\lambda+{1\over n}\right)\delta_{\sigma\sigma^\prime},
\end{eqnarray}
where $\rho_0 = N/L$. We can now just read off the velocities,
\begin{eqnarray}
v_s&=&v^s_N = v^s_J = {\pi\hbar\rho_0\over m}\left(\lambda+{1\over n}\right)\\
v^c_N&=&{\pi\hbar\rho_0\over m}n\left(\lambda+{1\over n}\right)^2\\
v^c_J&=&{\pi\hbar\rho_0\over m}{1\over n}\\
v_c&=&v_s
\end{eqnarray}
The charge and spin velocities are same for all $\lambda$ and
$n$. As expected from the singlet nature of the ground state the
ratio $v^s_J/v^s_N$ does not get renormalized.
The coefficient $v^c_J$ is independent of the interaction term, as a
consequence of Galilean invariance, but $v^c_N$ gets renormalized due to
non-linear interaction terms not included in the linear form of the harmonic
Hamiltonian.
The renormalization
coupling constant for the charge is
$\exp(-2\varphi)\rightarrow (v^c_N/v^c_J)^
{1/2}=n\lambda+1$ in the limit of long wavelength (i.e. $k\rightarrow 0$).
Because of the scale invariance of the model the dimensionless coupling
constant $\varphi$ is independent of the particle density.

The compressibility per particle is $(\pi^2\hbar^2\rho^2_0/m)(\lambda +
1/n)^2$.
We also find the chemical potential and the ground state energy
to be $m^*v_F^2/2$ and $
(N/3)(m^*v_F^2/2)$, respectively.  The chemical potential (or the
Fermi energy at zero temperature) and
the ground state energy of this one-dimensional system is that
of free fermions with the renormalized mass per particle
$m^*=m/(n\lambda+1)^2$.  The Fermi velocity is given by $v_F = \pi\hbar\rho_0
/(nm^*)$.

The low energy excitation of the lattice version is given by Eq. (\ref{lowe})
as well if we set $m=\hbar=1$ and replace $L$, $e_0$ and
$\mu$ with $N_a$, $-\pi^2/3+2e_0$, and $-\pi^2/3+2\mu$, respectively.
The crucial difference is that this expression for the lattice model is true
only for $\rho_0(=N/N_a) \le \rho_0^{max}(=(l+1/n)^{-1})$.
Hence, the charge velocity is
linear in $\rho_0$ only up to $\rho_0^{max}$.
We expect $v_c$ to vanish as $\rho_0\rightarrow 1$ due to
the metal-insulator transition at the density.
$v_c$, therefore, should exhibit non-analyticity
at $\rho_0^{max}$.  This behavior is attributed to the long-range
interaction in our model.
In ref. \cite{duncan2} this type
of non-analyticity in the spinon velocity was observed for the
SU(2) spin chain model.
In the nearest neighbor interaction models the sharp
change in the charge velocity is smoothed out.

For $l=1$  the energy for $\rho_0 \ge (1+1/n)^{-1}$ is explicitly found
and the harmonic fluid Hamiltonian is given by Eq. (\ref{lowe}) with
$A_{\sigma\sigma^\prime}
= (1-\rho_0)(n+2)-\rho_0/n + \delta_{\sigma\sigma^\prime}$ and
$B_{\sigma\sigma^\prime} = -\rho_0/n + \delta_{\sigma\sigma^\prime}$.
Therefore, $v_c = \pi(1-\rho_0)(n-1)$ and $v_s = \pi$.
The spin velocity is independent of the density in this region.
Similarly, for $l=0$ we have $A_{\sigma\sigma^\prime}
=B_{\sigma\sigma^\prime}=-\rho_0/n+\delta_{\sigma\sigma^\prime}$
for $0\le \rho_0\le 1$.
Thus, $v_c=\pi(1-\rho_0)$ and $v_s=\pi$.
This agrees with the results obtained by Kuramoto and Yokoyama
\cite{ky}.  However, it is unexpected that
the charge and spin
velocities are independent of the number of spin species.
Fig. 4 shows the charge and spin velocities for the lattice models
with various interaction parameters.

\section{Conclusion}
We have shown in this paper that the Sutherland's model of spinless
particles interacting with inverse square exchange can be generalized to
the multicomponent system of particles.  We further generalized the
continuum model to lattice for the integer values of the dimensionless
interaction parameter.  We explicitly constructed  a class of eigenstates for
these models and calculated the corresponding eigenenergies.
We claimed in this article that the class of eigenstates
correspond to the ground state and the low energy excitations
of the model.  We checked this by
the exact numerical diagonalization of the lattice version of
Hamiltonian for small systems.
The systematic construction of the other eigenstates, thermodynamics,
and the general n-point correlation functions
for these models are yet to be found.
As J. A. Minahan showed in his
preprint \cite{jam} the Hamiltonian of the type given by Eq. (\ref{firsteq})
also arises in a matrix model as a representation of one dimensional open
string theory.

This work is supported in part by NSF grant No. DMR-91-96212.

{\it Note added:} After the completion of this paper we received a preprint
from N. Kawakami who reports similar results.
And, the Hamiltonian given by Eq. (\ref{firsteq}) is also independently
identified by A. P. Polychronakos \cite{poly}.

\appendix{Proof of {\it Theorem 1}}
We first
discuss a diagrammatic way of writing the exchange operation.
A labeled diagram is shown in Fig.~\ref{diag}.
The amplitude of the labeled diagram shown in Fig.~\ref{diag}
can be evaluated using
the following rules: i) the dashed line connecting the indices $i$ and
$j$ gets a factor $z_iz_j(z_i-z_j)^{-2} (1-\delta_{\sigma
,\sigma^\prime})\delta_{\sigma\sigma_i}\delta_{\sigma\sigma_j}$,
ii) each solid line connecting the
indices $i$ and $\alpha_k$ gets a factor $(z_i-z_j)(z_{\alpha_k}
-z_i)^{-1}\delta_{\sigma\sigma_{\alpha_k}}$,
iii) each solid line connecting the indices
$j$ and $\beta_k$ gets a factor $-(z_i-z_j)(z_{\beta_k}-z_j)^{-1}
\delta_{\sigma\sigma_{\beta_k}}$.  Weight of the dashed line
vanishes whenever the spins at the sites $i$ and $j$ are same.
On the other hand, weights of the solid lines that are connected
to $i$($j$) vanish whenever the spins at the sites $\{\alpha\}$
($\{\beta\}$) and $i$($j$) are different.
Hence, we may think of the diagram as two-spin
interaction diagram. Since the Hamiltonian
has only the two-spin exchange operator $P_{ij}$, three-spin and
higher number of spin interactions are missing.

The amplitude of the labeled diagram is undesirable
since it depends on the local variables.
A more desirable, local variable independent diagram (unlabeled
diagram)
can be obtained by summing
$(n+1)(m+1)$ cyclically permuted labeled diagrams.
This fact will be proved
in this appendix.

We note that the indices, $\{i,\alpha_1,\ldots,\alpha_n\}$ and
$\{j,\beta_1,\ldots,\beta_m\}$,
are dummy indices. Consequently, the sum will be invariant under any
permutation of the indices.
We further observe that $\Delta$ is invariant under any permutation of the
indices except when there is interchange
between the two sets, $\{i,\alpha_1,\ldots,\alpha_n\}$ and $\{j,\beta_1,
\ldots,\beta_m\}$.  On the other hand,
the pre-factor to $\Delta$ remains invariant with respect to the permutations
of indices $\{\alpha\}$ and $\{\beta\}$.  It is
convenient to consider only the cyclic permutations which give $(n+1)(m+1)$
distinct pre-factors. We, then, sum the $(n+1)(m+1)$ factors as follow:

\begin{eqnarray}
P_{nm}^{\sigma\sigma^\prime}&=&\sum_{i\ne j}\sum_{\{\alpha\},\{\beta\}}{(-1)^m
\over n!\,m!}{z_i z_j^{1-q} (z_i-z_j)^{n+m-2+q}\over
(z_i-z_{\alpha_1})\cdots(z_i-z_{\alpha_n})
(z_j-z_{\beta_1})\cdots(z_j-z_{\beta_m})}\Delta \nonumber \\
&=&\sum_{i\ne j}\sum_{\{\alpha\},\{\beta\}} \sum_{k,l} {(-1)^m \over
(n+1)!\,(m+1)!}{z_{\gamma^\alpha_k} z_{\gamma^\beta_l}^{1-q}
(z_{\gamma^\alpha_k}-z_{\gamma^\beta_l})^{n+m-2+q}\over
(z_{\gamma^\alpha_k}-z_{\alpha_1})\cdots(z_{\gamma^\alpha_k}-z_i)\cdots
(z_{\gamma^\alpha_k}-z_{\alpha_n})} \nonumber \\
&\times&{1\over (z_{\gamma^\beta_l}-z_{\beta_1})\cdots(z_{\gamma^\beta_l}-z_j)
\cdots (z_{\gamma^\beta_l}-z_{\beta_m})}\Delta ,\label{a1}
\end{eqnarray}

where $\{\gamma^\alpha\}\equiv \{i,\alpha_1,\ldots,\alpha_n\}$ and
$\{\gamma^\beta\}\equiv \{j,\beta_1,\ldots,\beta_m\}$.

We let $D_\alpha=\prod_{k<k^\prime}(z_{\gamma^\alpha_k}-
z_{\gamma^\alpha_{k^\prime}})$ and $D_\beta=\prod_{l<l^\prime}
(z_{\gamma^\beta_l} - z_{\gamma^\beta_{l^\prime}})$.  Using the binomial
theorem we expand the numerator in Eq.\ (\ref{a1}).  We then multiply and
divide the expression by $D_\alpha D_\beta$ and obtain the following.
\begin{equation}
P^{\sigma\sigma^\prime}_{nm}=\sum_{i\ne
j}\sum_{\{\alpha\},\{\beta\}}\sum_{s=1}^{
n+m-1+q}{(-1)^{m-1+q-s}\over (n+1)!\,(m+1)!}{n+m-2+q \choose s-1}{V^s_\alpha
V^{n+m-s}_\beta\over D_\alpha D_\beta} \Delta, \label{vand}
\end{equation}
where
\begin{equation}
V^s_\alpha=\left|\matrix{
1&1\hfill&\ldots&1\hfill\cr
z_i&z_{\alpha_1}\hfill&\ldots&z_{\alpha_n}\hfill\cr
\,\vdots\hfill&\,\vdots\hfill&&\,\vdots\hfill\cr
z_i^{n-1}&z_{\alpha_1}^{n-1}\hfill&\ldots&z_{\alpha_n}^{n-1}\hfill\cr
z_i^s&z_{\alpha_1}^s\hfill&\ldots&z_{\alpha_n}^s\hfill\cr}\right|,
\end{equation}
and $V^k_\beta$ is defined the same way.  We further note that
$V^s_\alpha = 0$ if $0\le s\le n-1$ and $V^k_\beta = 0$ if $0\le k\le m-1$.
Therefore, if
$0\le q \le 1$, the only non-zero contribution in (\ref{vand}) is the term with
$s=n$. Since $V^n_\alpha = D_\alpha$ and $V^m_\beta = D_\beta$, we obtain
\begin{eqnarray}
P^{\sigma\sigma^\prime}_{nm}&=&{(-1)^{n+m-1+q}\over
(n+1)!\,(m+1)!}{n+m-2+q\choose n-1}\sum_{i\ne
j}\sum_{\{\alpha\},\{\beta\}}\Delta \nonumber \\
&=&(-1)^{n+m-1+q}{n+m-2+q\choose n-1}
{M_\sigma\choose n+1}{M_{\sigma^\prime}\choose m+1}.\label{sumd}
\end{eqnarray}
Evaluating $\sum_{i\ne j}\sum_{\{\alpha\},\{\beta\}}\Delta$ in (\ref{sumd}) is
same as calculating the total number of ways of putting $n+1$ out of $M_\sigma$
blue balls in a box and $m+1$ out of $M_{\sigma^\prime}$ red balls in
another box.
The sum, therefore, is equal to $M_\sigma(M_\sigma-1)\cdots (M_\sigma-n)M_
{\sigma^\prime}(M_{\sigma^\prime}-1)\cdots(M_{\sigma^\prime}-m)$.

It is straightforward to prove the following useful identities:
\begin{eqnarray}
{n+m-2+q\choose n-1}&=&\sum_{s=0}{n-1\choose s}{m-1+q\choose s},\label{sep}\\
(-1)^s{m-1+q \choose s}&=&\sum_{k=0}^{s}(-1)^k{m+q\choose k},\nonumber \\
&=&\sum_{k=0}^{s}(-1)^k(s+1-k){m+1+q\choose k},\label{sep2}
\end{eqnarray}
\begin{equation}
\sum_{m=1-q}^{M_\sigma-1}(-1)^m{M_\sigma\choose m+1}{m+1\choose k}=\left\{
\begin{array}{l}
\left.
\matrix{
-M_\sigma+1&\mbox{if}\hfill&k=0\hfill\cr
-M_\sigma&\mbox{if}\hfill&k=1\hfill\cr
0&\mbox{for}\hfill&2\le k < M_\sigma\cr}
\right\}\;\;\mbox{for}\;q=0,\\
\left.
\matrix{
1&\mbox{for}\hfill&k=0\hfill\cr
0&\mbox{for}\hfill&1\le k < M_\sigma\hfill\cr}
\right\}\;\;\mbox{for}\;q=1,
\end{array}\right. \label{sep3} 
\end{equation}
Using (\ref{sep}) we may sum the terms
depending on $n$ and $m$ in $P^{\sigma\sigma^\prime}_{nm}$
separately. Using (\ref{sep2}) and (\ref{sep3})
the following relation can easily be obtained:
\begin{equation}
P^{\sigma\sigma^\prime}=\left\{
\begin{array}{l}
-\sum_{k=1}^{Min(M_\sigma,M_{\sigma^\prime})}(M_\sigma-k)(M_{\sigma^\prime}-k)
 \;\;\mbox{for}\; q=0\;,\\
-\sum_{k=1}^{Min(M_\sigma,M_{\sigma^\prime})}(M_\sigma-k)\;\; \mbox{for}\;
q=1\;.
\end{array}\right.\nonumber 
\end{equation}
{\hfill Q.E.D.}
\appendix{Proof of {\it Theorem 2}}
We replace the product over $k$ with two new sets of integer indices,
$\{\alpha\}$ and $\{\beta\}$ and
rewrite the expression in {\it Theorem 2} as,
\begin{eqnarray}
Q&=&\sum_{i\ne j}\sum_{\{\alpha\},\{\beta\}}{1\over (M_\sigma-1)!\,
(M_{\sigma^\prime}-1)!}{z_i\over z_j}\left({z_i-z_j\over z_j}\right)^{q-2}
{z_{\alpha_1}-z_j\over z_{\alpha_1}-z_i}
{z_{\alpha_2}-z_j\over z_{\alpha_2}-z_i}
\cdots {z_{\alpha_{M_\sigma-1}}-z_j\over z_{\alpha_{M_\sigma-1}}-z_i}
\nonumber \\
&\times&\left\{\sum_{r=0}^{M_{\sigma^\prime}-1}(z_i-z_j)^r
f_r(\{z_\beta\})\right\}\Delta.
\label{exp}
\end{eqnarray}
Here, $\Delta$ is same as in {\it Theorem 1\/}.
The factor in the curly bracket in (\ref{exp}) is obtained by writing the terms
like $(z_{\beta_l}-z_i)/(z_{\beta_l}-z_j)$ as
$\{1-(z_i-z_j)/(z_{\beta_l}-z_j)\}$ and by multiplying out all the factors that
depend on $\{\beta\}$.
$f_r$ in (\ref{exp}) is some function of $\{z_\beta\}$, however, $f_0=1$.

As in Appendix A we rewrite the expression by summing over the factors obtained
by the cyclic permutations of the indices
$\{i,\alpha_1,\ldots,\alpha_{M_\sigma-1}\}$ and obtain
\begin{equation}
Q=\sum_{i\ne j}\sum_{\{\alpha\},\{\beta\}}\sum_{r=0}^{M_{\sigma^\prime}-1}
{(-1)^{M_\sigma}\over M_\sigma !\,
(M_{\sigma^\prime}-1)!}(z_i-z_j)(z_{\alpha_1}-z_j)\cdots
(z_{\alpha_{M_\sigma}-1} -z_j)f_r{W^{M_\sigma}_\alpha\over
D^{M_\sigma}_\alpha},
\end{equation}
where $W^{M_\sigma}_\alpha$ is given by the Vandemonde determinant whose last
row is modified to $z_{\gamma_l}(z_{\gamma_l}-z_j)^
{q+r-3}/z_j^{q-1}$, $\{\gamma_l\}\equiv \{i,
\alpha_1,\ldots,\alpha_{M_\sigma-1}\}$.
It is straightforward to show that
\begin{eqnarray}
W^{M_\sigma}_\alpha&=&\left\{
\begin{array}{l}
{(-1)^{M_\sigma} D^{M_\sigma}_\alpha\over (z_i-z_j)(z_{\alpha_1}-z_j)\cdots
(z_{\alpha_{M_\sigma}-1} -z_j)}
\;\;\mbox{if}\;q=2\;\mbox{and}\;r=0\\
0\;\;\mbox{if}\;0\le q+r-3\le M_\sigma-3
\end{array}\right.,\\
\sum_{i\ne j}\sum_{\{\alpha\},\{\beta\}}\Delta&=&M_\sigma !\,M_{\sigma^\prime}!
\end{eqnarray}
The only non-zero contribution to $Q$ is given by the condition $q=2$ and
$r=0$.  Since $0\le r\le M_{\sigma^\prime}-1$, the sufficient condition for
$W^{M_\sigma}_\alpha=0$ is $3\le q\le M_\sigma - M_{\sigma^\prime}+1$.
Therefore, we have for $M_\sigma \ge M_{\sigma^\prime}$
\begin{equation}
Q=\left\{
\begin{array}{l}
M_{\sigma^\prime}\;\; \mbox{for}\; q=2,\\
0 \;\;\mbox{for} \;3 \le q \le M_\sigma-M_{\sigma^\prime}+1.
\end{array}\right. 
\end{equation}

{\hfill Q.E.D.}

\figure{Evolution of the inverse square exchange models both in the continuum
space and lattice.  Models that already exist are in rectangular boxes and new
models that are presented in this paper are in ovals.  The models enclosed in
the dotted box are continuum version and the rest of the models are lattice
version.  The solid arrows mean ``evolved from'' or ``generalized from''.
The dashed  arrows connect two equivalent models in disguise.
\label{origin}}

\figure{Labeled diagram for $n$ $\sigma$-spins and $m$ $\sigma^\prime$-spins.
\label{diag}}

\figure{Illustration of allowed values of the currents for the
sector $\{14,12,10,6,4,2\}$.  (a) Allowed. All
currents are zero. (b) Allowed. Ground state for the sector.
(c) Not allowed.
\label{curr}}

\figure{Charge and spin velocities of the lattice models.  (a) Fermionic
supersymmetric t-J model.  The velocities are independent of the number of
spin species.  (b) Bosonic supersymmetric t-J model.  Up to the density
$n/(1+n)$, $v_c$ and $v_s$ are equal.  (c) Bosonic and fermionic t-J models
with $l\ge 2$.  The charge and spin velocities are same up to the density
$(l+1/n)^{-1}$.  Beyond the density the ground state is no longer the
Jastrow-product type.
\label{velocities}}

\vfill\eject
fig.1
\vfill\eject
\begin{picture}(100,200)(10,20)
\put(30,210){Fig. 2}
\multiput(180,0)(15,0){10}{\line(1,0){8}}
\put(177,-2){$\bullet$} \put(320,-2){$\bullet$}
\put(180,0){\line(-1,1){100}}
\put(322,0){\line(1,1){100}}
\put(180,0){\line(-1,-1){100}}
\put(322,0){\line(1,-1){100}}
\put(180,0){\line(-2,1){100}}
\put(322,0){\line(2,1){100}}
\multiput(80,25)(0,-25){6}{.}
\multiput(420,25)(0,-25){6}{.}
\put(77,97){$\bullet$} \put(77,47){$\bullet$} \put(77,-103){$\bullet$}
\put(420,97){$\bullet$} \put(420,47){$\bullet$} \put(420,-103){$\bullet$}
\put(178,5){$i$} \put(319,5){$j$}
\put(65,97){$\alpha_1$} \put(65,47){$\alpha_2$} \put(65,-103){$\alpha_n$}
\put(424,97){$\beta_1$} \put(424,47){$\beta_2$} \put(424,-103){$\beta_m$}
\end{picture}
\vfill\eject
\begin{picture}(100,200)(10,20)
\put(30,210){Fig. 3}
\put(0,-10){(a)}
\multiput(0,150)(30,0){14}{\framebox(30,30){}}
\multiput(0,120)(30,0){12}{\framebox(30,30){}}
\multiput(0,90)(30,0){10}{\framebox(30,30){}}
\multiput(0,60)(30,0){6}{\framebox(30,30){}}
\multiput(0,30)(30,0){4}{\framebox(30,30){}}
\multiput(0,0)(30,0){2}{\framebox(30,30){}}
\put(0,-210){(b)}
\multiput(0,-50)(30,0){14}{\framebox(30,30){}}
\multiput(30,-80)(30,0){12}{\framebox(30,30){}}
\multiput(60,-110)(30,0){10}{\framebox(30,30){}}
\multiput(120,-140)(30,0){6}{\framebox(30,30){}}
\multiput(150,-170)(30,0){4}{\framebox(30,30){}}
\multiput(180,-200)(30,0){2}{\framebox(30,30){}}

\put(0,-410){(c)}
\multiput(0,-250)(30,0){14}{\framebox(30,30){}}
\multiput(120,-280)(30,0){12}{\framebox(30,30){}}
\multiput(90,-310)(30,0){10}{\framebox(30,30){}}
\multiput(120,-340)(30,0){6}{\framebox(30,30){}}
\multiput(150,-370)(30,0){4}{\framebox(30,30){}}
\multiput(180,-400)(30,0){2}{\framebox(30,30){}}

\end{picture}
\vfill\eject
fig. 4
\vfill\eject

\begin{references}
\bibitem{suther} B. Sutherland, Phys. Rev. \ A \ {\bf 4}, 2019 (1971);
Phys. \ Rev. \ A \ {\bf 5}, 1372 (1972).
\bibitem{hasha} F. D. M. Haldane,
Phys. \ Rev. \ Lett. \ {\bf 60}, 635 (1988); B. Sriram Shastry,
Phys. \ Rev. \ Lett. \ {\bf 60}, 639 (1988).
\bibitem{duncan} F. D. M. Haldane,
Phys. \ Rev. \ Lett. \ {\bf 47}, 1840 (1982); {\bf 48}, 569(E) (1982).
\bibitem{duncan2} F. D. M. Haldane,
Phys. \ Rev. \ Lett. \ {\bf 66}, 1529 (1991).
\bibitem{ky} Y. Kuramoto and H. Yokoyama,
Phys. \ Rev. \ Lett. \ {\bf 67}, 1338 (1991).
\bibitem{poly} A. P. Polychronakos, preprint CU-TP-551.
\bibitem{jam} J. A. Minahan, preprint UVA-HET-92-01.
\end{references}
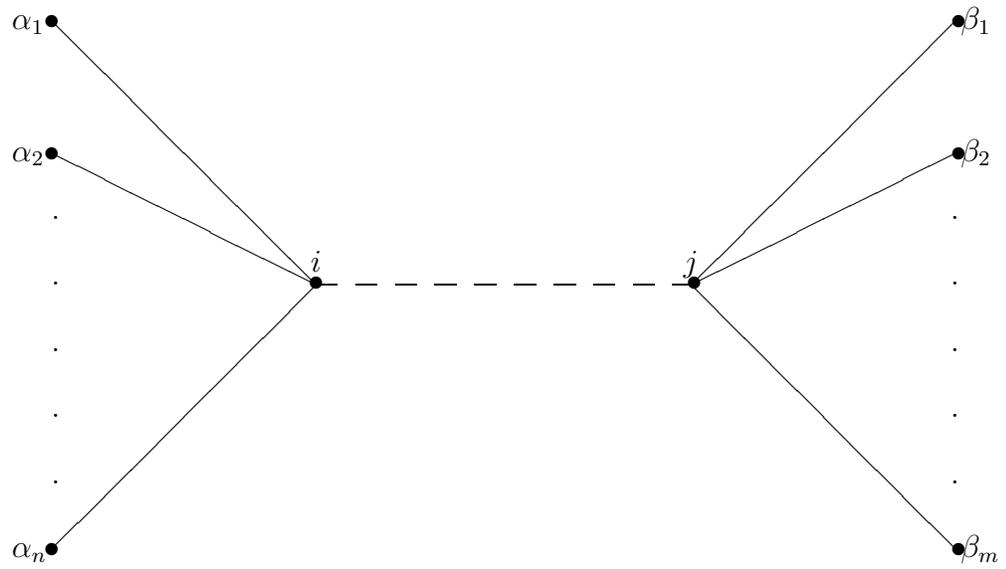
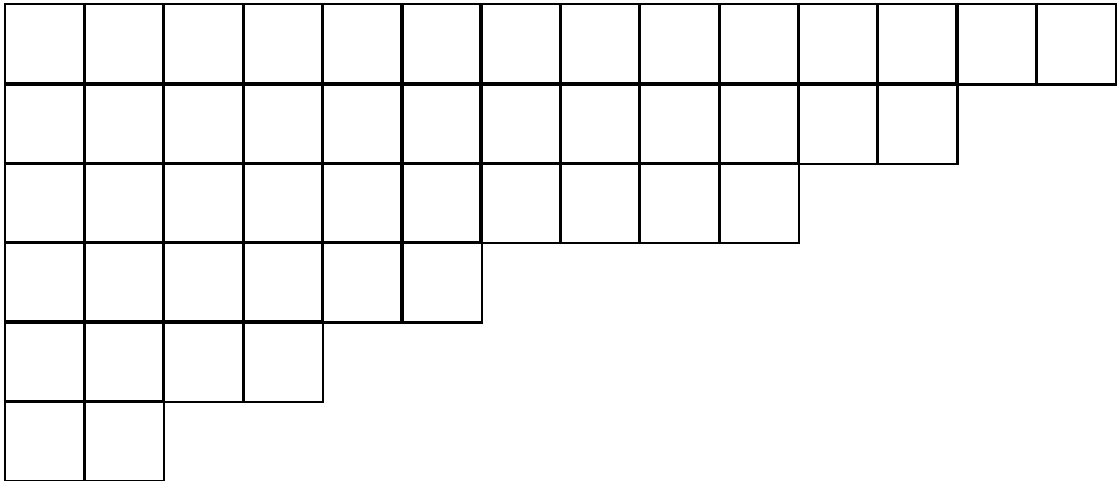
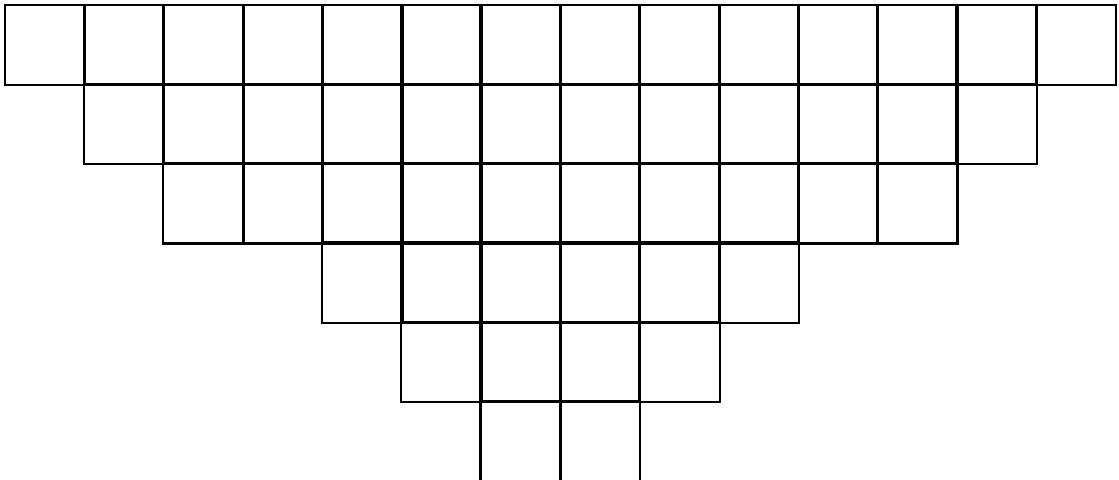
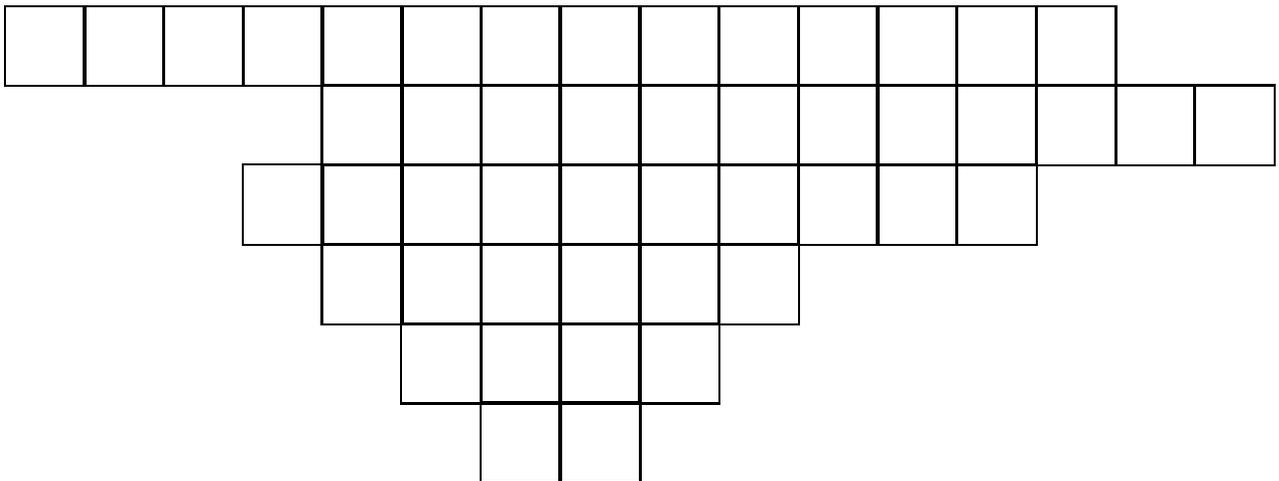
\end{document}